\documentclass{article}

\usepackage{arxiv}

\usepackage[utf8]{inputenc} 
\usepackage[T1]{fontenc}    
\usepackage{hyperref}       
\usepackage{url}            
\usepackage{booktabs}       
\usepackage{amsfonts}       
\usepackage{microtype}      
\usepackage{amsmath}
\usepackage{graphicx}
\usepackage{subcaption}

\let\ACMmaketitle=\maketitle
\renewcommand{\maketitle}{\begingroup\let\footnote=\thanks \ACMmaketitle\endgroup}

\title{The resistance of an FPGA implementation of the Grasshopper block cipher to CPA attacks\footnote{This article was also presented by Alexander A. Istomin and Éric Filiol at \textit{RusCrypto 2019}}}

\author{
  Cédric Delaunay \\
  Laboratoire de Cryptologie et de Virologie Opérationnelles \\
  ESIEA Laval \& ENSTA Bretagne \\
  Laval, France \& Brest, France\\
  \texttt{cedric.delaunay@ensta-bretagne.org} \\
}

\begin{document}
\maketitle

\begin{abstract}
In this paper, we implement the Russian standard block cipher Grasshopper on Field-Programmable Gate Array (FPGA). We also study the Correlation Power Analysis attack, which is a special type of side-channel attack proposed by Brier \textit{et al} \cite{cpa_init}. To face this kind of attack, we propose a solution of software countermeasure, and we present the associated implementation of the Grasshopper algorithm. These two implementations are then compared to an AES-256 one. Finally, through the implementation of a CPA attack on an FPGA development board, we show that typical attack models that work on AES fail on Grasshopper implementations.
\end{abstract}

\vspace{0.5cm}

\keywords{Grasshopper \and Kuznyechik \and Symmetric Encryption \and Block cipher \and Field-Programmable Gate Array \and Correlation Power Analysis \and Side-Channel Attacks}

\vspace{0.5cm}

\section{Introduction}

Russian Federal Agency on Technical Regulating and Metrology approved in 2015 the new block cipher of Russian Federation, which is called \textit{Kuznyechik} (or Grasshopper in English), and that is standardised by GOST R 34.12-2015 \cite{rfc7801}. This symmetric block cipher algorithm is based on a substitution-permutation network explained in Section \ref{sec:pres}.

In this research, we propose to implement and secure the algorithm on FPGA (acronym of \textit{Field-programmable Gate Array}) in order to study the performance of this algorithm on this type of circuit. Performances will be compared with an implementation of the AES-256 (\textit{Advanced Encryption Standard} with 256-bit key) algorithm, which is an international standard in terms of symmetric encryption since 2010 \cite{iso18033}.

An FPGA is an integrated circuit that can be configured and programmed by the user of the circuit according to the functions he wants to perform. Although this technology was introduced in the 1980s (Xilinx was founded in 1984 and the first FPGA was commercialised in 1985 \cite{fpga_history}), it is still very popular because of its qualities, such as high performance, moderate price and relatively easy maintainability. The success of this technology is evidenced by the fact that Altera, one of the leading FPGA manufacturers, was acquired by Intel in 2015 for \$ 16.7 billion, making this acquisition the largest financial transaction ever carried out by Intel \cite{altera_Intel}.

Nevertheless, this technology is sensitive to side-channel attacks, and more particularly to Differential Power Analysis attacks. This kind of attack, presented in 1999 by Kocher \textit{et al. } \cite{dpa}, is based on the fact that the power consumed by a semiconductor or the electromagnetic leaks it emits are correlated with the operations it performs. Thus, it is possible to find information about the encryption, including information about the key. This kind of attack works very well on FPGA implementations of AES \cite{cpa_aes}, and the purpose of this study is also to study the resistance of the Grasshopper algorithm to "classic" CPA attacks by providing a secure implementation with an efficient counter-measure. 

This article is organised as follows. In section \ref{sec:sec1}, the Grasshopper algorithm will be presented and its performance on FPGA discussed. Then, in section \ref{sec:sec2}, we will focus on presenting the threats we face, the countermeasures that exist and what solution we have implemented to secure the algorithm. Section \ref{sec:sec3} presents the attacks that have been carried out and the results we have obtained.

\section{First implementation of Grasshopper algorithm}
\label{sec:sec1}

The purpose of this section is to present in a little more detail the operations performed by the Grasshopper algorithm, as well as the methodology used to provide a standard implementation and how the measurement of the algorithm's performance was carried out.

\subsection{Presentation of the algorithm}
\label{sec:pres}

Grasshopper is a 128-bit block cipher algorithm that uses a 256-bit key \cite{rfc7801}. This key is used to generate ten 128-bit subkeys according to a Feistel network. Encryption takes place over 10 rounds,in which three main functions are applied to the state\footnote{The name of the functions is taken from the standard}.

\subsubsection{The add round key operation}

An exclusive OR (noted XOR in the rest of this article) $ X[k]$ between the current state and the subkey $ k $.

\subsubsection{The non linear permutation}

Noted $ S $, where each byte of the state is substituted with the corresponding byte of following S-Box noted $ S' $ :

\begin{align*}
    S' = (&\phantom{{}={}} 252, 238, 221, 17, 207, 110, 49, 22, 251, 196, 250, 218, 35, 197, 4, 77, \\
    &\phantom{{}={}} 233, 119, 240, 219, 147, 46, 153, 186, 23, 54, 241, 187, 20, 205, 95, 193,\\
    &\phantom{{}={}} 249, 24, 101, 90, 226, 92, 239, 33, 129, 28, 60, 66, 139, 1, 142, 79,\\
    &\phantom{{}={}} 5, 132, 2, 174, 227, 106, 143, 160, 6, 11, 237, 152, 127, 212, 211, 31,\\
    &\phantom{{}={}} 235, 52, 44, 81, 234, 200, 72, 171, 242, 42, 104, 162, 253, 58, 206, 204,\\
    &\phantom{{}={}} 181, 112, 14, 86, 8, 12, 118, 18, 191, 114, 19, 71, 156, 183, 93, 135,\\
    &\phantom{{}={}} 21, 161, 150, 41, 16, 123, 154, 199, 243, 145, 120, 111, 157, 158, 178, 177,\\
    &\phantom{{}={}} 50, 117, 25, 61, 255, 53, 138, 126, 109, 84, 198, 128, 195, 189, 13, 87,\\
    &\phantom{{}={}} 223, 245, 36, 169, 62, 168, 67, 201, 215, 121, 214, 246, 124, 34, 185, 3,\\
    &\phantom{{}={}} 224, 15, 236, 222, 122, 148, 176, 188, 220, 232, 40, 80, 78, 51, 10, 74,\\
    &\phantom{{}={}} 167, 151, 96, 115, 30, 0, 98, 68, 26, 184, 56, 130, 100, 159, 38, 65, \\
    &\phantom{{}={}}  173, 69, 70, 146, 39, 94, 85, 47, 140, 163, 165, 125, 105, 213, 149, 59, \\
    &\phantom{{}={}} 7, 88, 179, 64, 134, 172, 29, 247, 48, 55, 107, 228, 136, 217, 231, 137,\\
    &\phantom{{}={}} 225, 27, 131, 73, 76, 63, 248, 254, 141, 83, 170, 144, 202, 216, 133, 97,\\
    &\phantom{{}={}} 32, 113, 103, 164, 45, 43, 9, 91, 203, 155, 37, 208, 190, 229, 108, 82,\\
    &\phantom{{}={}}89, 166, 116, 210, 230, 244, 180, 192, 209, 102, 175, 194, 57, 75, 99, 182)
\end{align*}

\newpage
\subsubsection{The linear transformation}

Noted $ L $, the linear function of Grasshopper is based on another function noted $ R $. Assuming that state $ x $ is noted $ x_{15} || x_{14} || ... || x_{0} $\footnote{$ || $ is the symbol for concatenation }, we have : 

\begin{align*}
    L(x) &= R^{16}(x) \\
    \intertext{where} R(x) &= l(x_{15}, ..., x_{0}) || x_{15} || x_{1} \\
    \intertext{with} l(x_{15}, ..., x_{0}) &= 148 x_{15} + 32 x_{14} + 133 x_{13} + 16 x_{12} \\
    &+  194 x_{11} + 192 x_{10} + 1 x_{9} + 251 x_{8} \\
	&+ 1 x_{7} + 192 x_{6} + 194 x_{5} + 16 x_{4} \\
	&+ 133 x_{3} + 32 x_{2} + 148 x_{1} + 1 x_{0} \\
	&\phantom{{}={}} \\
	\intertext{where all multiplications are performed in}  GF(2)[X] &/ p(x)\\
	\intertext{with} p(x) = x^{8} + x^{7} &+ x^{6} + x + 1  \in GF(2)[X]
\end{align*}

If $ P $ is the plaintext, and $ C $ the resulting ciphertext, then the encryption routine is written as follows (or can be represented as the diagram shown in Figure \ref{fig:fig1}):
\begin{align*}
C &= X[k_{10}] L S X[k_9] L S X[k_8] ... L S X[k_2] L S  X[k_1](P)
\end{align*}

\vspace{0.5cm}

\begin{center}
  \captionsetup{type=figure}
  \includegraphics[width=16cm]{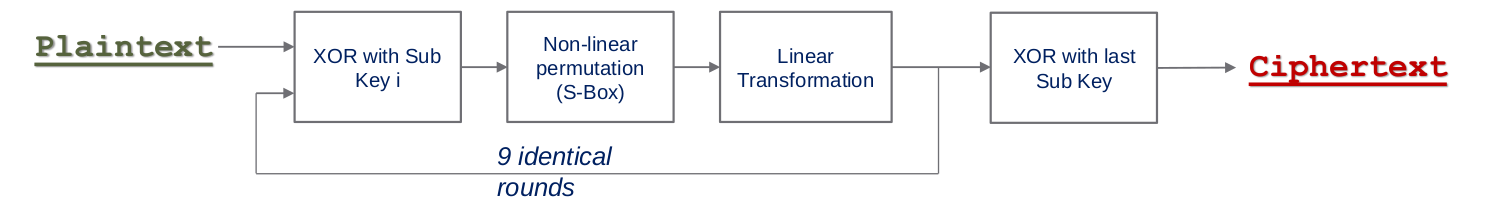}
  \captionof{figure}{Operating diagram of Grasshopper's encryption routine}
  \label{fig:fig1}
\end{center}


\subsection{First implementations and performances}

The first objective of this research was to provide an optimised implementation of the Grasshopper algorithm. Thus, several versions of the algorithm have been developed, in order to compare their performances, both in terms of encryption rate and compactness. For example, the fastest version is the one where the ten rounds of the algorithm are performed on a single clock cycle, but it is also the least compact.

The results of these implementations can be observed on the graphs below representing the encryption delay and encryption speed respectively depending on the compactness of the model. 

\begin{center}
  \captionsetup{type=figure}
  \includegraphics[width=16cm]{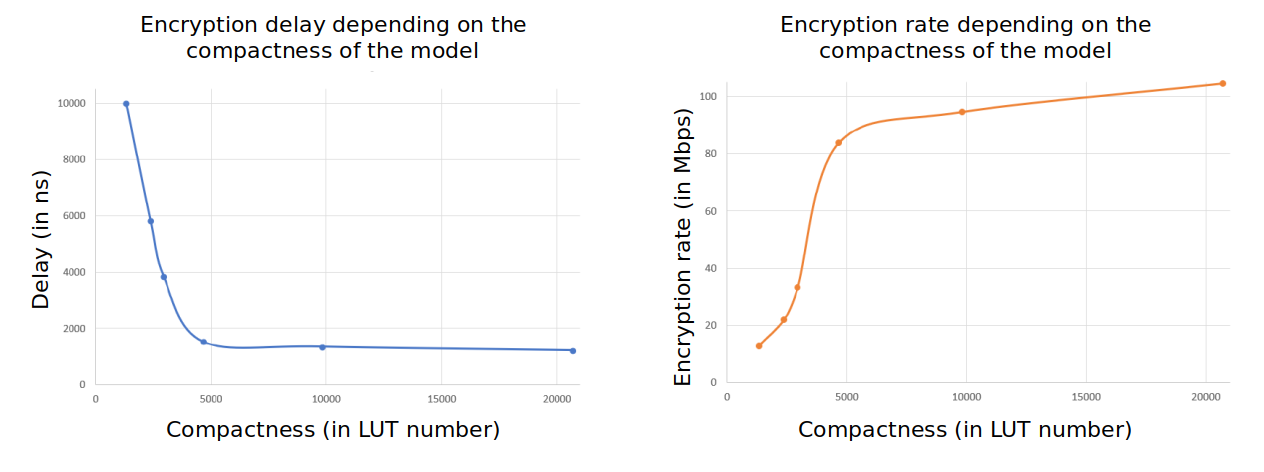}
  \captionof{figure}{Diagrams of encryption delay and encryption rate of Grasshopper FPGA implementations depending on the Compactness of the design}
  \label{fig:fig2}
\end{center}

\section{Securing the algorithm against DPA attacks}
\label{sec:sec2}

There are two main types of attacks: physical attacks and software attacks. In this article, we study the resistance of the Grasshopper algorithm to CPA attacks, which are part of the first category. As you can see in Figure \ref{fig:fig3}, physical attacks can be divided into two sub-categories, namely invasive attacks (which involve the integrity of the attacked component) and non-invasive attacks. This sub-category includes active and passive attacks, which include CPA attacks, which are side-channel attacks.

\begin{center}
  \captionsetup{type=figure}
  \includegraphics[width=10cm]{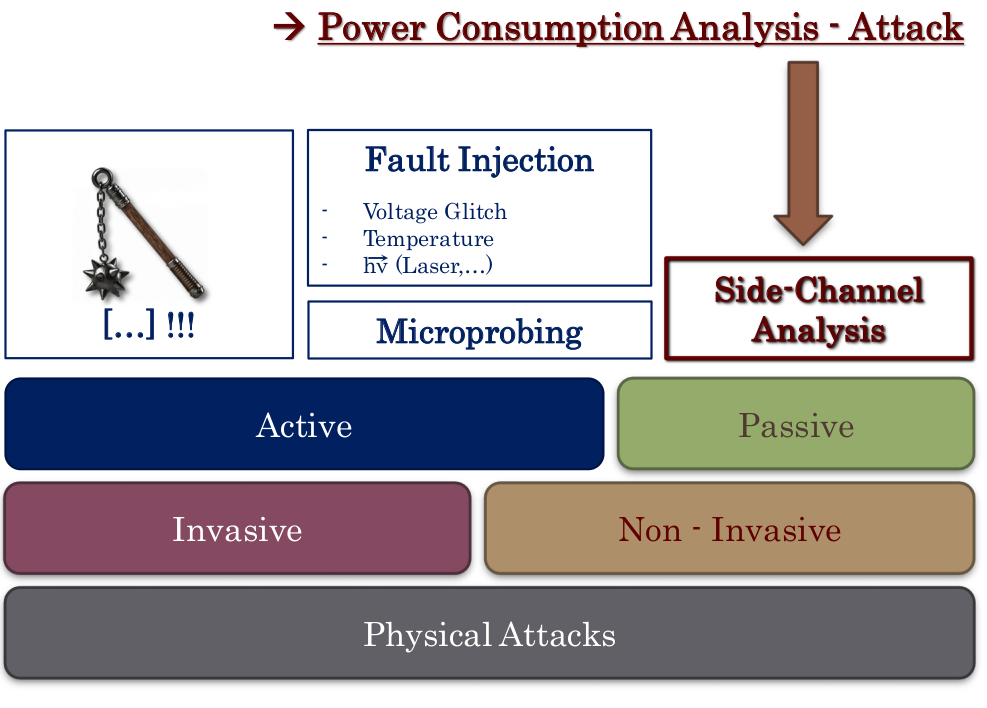}
  \captionof{figure}{Schema of the different categories of physical attacks}
  \label{fig:fig3}
\end{center}

\subsection{The concept of a CPA attack}

For a detailed description of DPA / CPA attack, the reader is invited to refer to \textit{Correlation Power Analysis with a Leakage Model} \cite{cpa_init}, \textit{Hardware Security: Design, Threats, and Safeguards}\cite{hardwareSec} and \textit{Security of Block Ciphers: From Algorithm Design to Hardware Implementation} \cite{securityBC}.

A CPA attack has several phases. First, the attacker will acquire a large number of traces, by encrypting a number N of plaintexts (and recovering their associated ciphertext). Most of the time, these traces are acquired thanks to a field probe (electromagnetic leaks) or a differential probe (power consumed).

Once the traces have been recovered, it is necessary to have "reference" data to compare our traces against, in order to deduce information about the key. This is done by establishing a leakage model, which describes the theoretical evolution of the leaks of our circuit. The most common leakage models are the Hamming distance and the Hamming weight, but there are others, such as the single-bit model \cite{securityBC}. In the figure \ref{fig:fig4}, we can see that the voltage of the tested circuit depends directly on the Hamming Weight or Hamming Distance of the processed data.

\begin{center}
  \captionsetup{type=figure}
  \includegraphics[width=10cm]{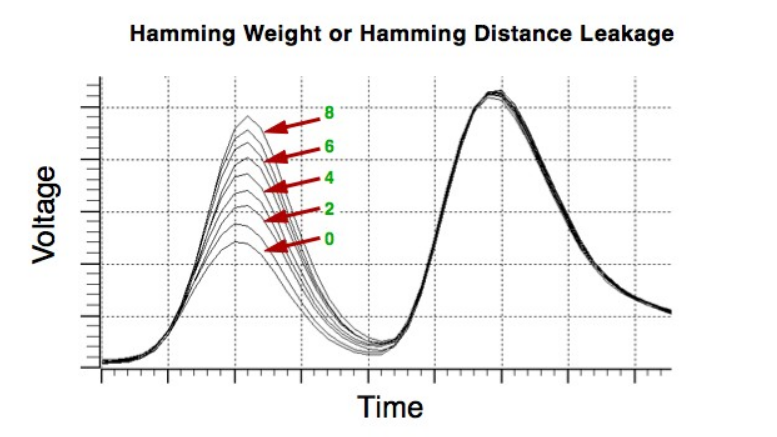}
  \captionof{figure}{Hamming Weight or Hamming Distance leakage \cite{hd}}
  \label{fig:fig4}
\end{center}

Now that the leakage model is established, it is necessary to take the pairs of plaintext / ciphertext from the previous step and pass them in our leakage model. Here, let us take the case of a classic CPA attack on AES, which exploits a vulnerability of the last turn \cite{aes_cpa_newae}. The idea will be to apply the reversed last round of AES on all the ciphertexts as well as our leakage model, and this for each possible value of the subkey.

Now, real leaks and theoretical leaks have been obtained so they need to be compared. It is then necessary to use a mathematical distinguisher, such as the Pearson correlation coefficient \cite{pearson} for example, in order to find the value of the most likely subkey.

\subsection{The different types of countermeasures}

In order to prevent attackers from performing DPA attacks, two main types of countermeasures exist : the hardware ones and the software ones. Here are some of the main hardware solutions :
\begin{itemize}
\item The noise generator \cite{ng}. Indeed, by generating noise on the power consumed by our FPGA, we can hide the encryption steps and it becomes more complicated to accurately distinguish the steps from the encryption. A similar technique consists in generating a signal of maximum power at any time during encryption. However, this technique is no longer, or rarely used, because it is very demanding in terms of power and is therefore not suitable for a battery-powered system.
\item the insertion of a random delay during encryption steps \cite{rdi}. This method consists of randomly inserting "dummy" instructions into the encryption that are not useful, but will change the encryption delay. The traces resulting from this process will therefore all be different and more difficult to interpret
\item the Shuffling \cite{shuf}. This technique is based on the fact that some steps of an encryption algorithm can be performed independently of their order. Thus, by mixing the execution order of certain functions (such as the shift of each byte in the S-Box), we will introduce a notion of uncertainty and each execution of an encryption routine will be "different" from the previous one.
\end{itemize}

In addition to these hardware countermeasures, there are various software countermeasures. Finally, we will discuss about \textit{Dual-rail with Precharge Logic} (DPL) \cite{dpl} and \textit{masking}, and why the masking solution was chosen. First, DPL consists in coding each bit with two bits. For instance, "0" is coded "01" and "1" is coded "10". The aim of this technique is to "homogenise" the power consumption of the circuit, because sometimes, the power consumed or the electromagnetic leaks can depend on the number of "1" in a data bus, or the number of "0" to "1" transitions of a variable. With DPL, the variables always has the same number of "0" and the same number of "1". Actually, DPL makes it possible to remove the bias present in the information on which the processor is working, a bit in the style of the Von Neumann corrector \cite{vonNeumann} (the Von Neumann corrector is a data compression method that removes the bias initially present in the data stream). 

The other software main countermeasure is the masking method \cite{mask}. It consists in XORing the plaintext with a mask before the encryption in order to realise the encryption routine on a state that is not directly correlated with the plaintext/ciphertext couple. Thus, the leaks emitted by the circuit will not be correlated with this couple neither.

As a result, when it came to choosing a software countermeasure, masking quickly became a must. Indeed, unlike DPL, masking is lighter, and does not require rewriting the entire design. Indeed, the major disadvantage of DPL is that without countermeasure, the states on which the algorithm works are 128 bits, and should then be 256 bits with DPL.

\subsection{Masking the algorithm}

In practice, here is how masking works. Let $ x $ be the plaintext and $ m $ the mask. Before encryption, an exclusive OR is performed between $ x $ and $ m $, noted $ + $. For $ X $ and $ L $ functions, $ X[k](x+m) = X[k](x) + m $ and $ L(x + m) = L(x) + L(m) $ because $ X $ and $ L $ are linear applications. For $ S $ on the other hand, it is a little more delicate, because this permutation is non-linear. To avoid this slight problem, an application $ S_m $ is built, also non-linear, such as $ S_{m}(x + m) = S(x) + m $ and this permutation $ S_m $ will be used in the encryption routine instead of the standard S-Box.

So, the first round of the Grasshopper algorithm gives:
\begin{align*} LS_{m1}X[k_1](x+m) &= LS_{m1}(X[k_1](x) + m) \\
								&= L(S(X[k_1](x)) + m)) \\
								&= LSX[k_1](x) + L(m) \\
\end{align*}

In order to have a fully masked encryption, this round needs to be repeated nine times, and followed by the XOR with the last subkey:
\begin{align*} 
X[k_{10}]LS_{m9}X[k_9]...LS_{m1}X[k_1](x + m) = X[k_{10}]LSX[k_9]...LSX[k_1](x) + L^{9}(m)
\end{align*}

To get the ciphertext, the resulting state needs to be XORed with $ L^9 (m) $.

\subsection{Performances of the masked implementation}

\begin{center}
\begin{tabular}{|c|c|c|c|}
\hline
Algorithm & Standard Grasshopper & Masked Grasshopper & AES-256 \\\hline
Look-Up Tables & 8810 & 10106 & 4846 \\\hline
Latches & 2167 & 2697 & 4540 \\\hline
Operating Frequency & 28,5 MHz & 28,5 MHz & 28,5 MHz \\\hline
Delay & 1526,2 ns & 1596,2 ns & 1920 ns \\\hline
Encryption Rate & 83,9 Mbps & 80,8 Mbps & 66,7 Mbps\\\hline
\end{tabular}
\captionsetup{type=table}
\captionof{table}{Comparative table between standard Grasshopper implementation, Masked Grasshopper Implementation and AES-256}
\end{center}

First of all, by looking at the table, we notice that the secure implementation requires a few more elements than the classic implementation: 30\% more Look-up tables and 20\% more latches. These numbers are the consequence of the implementation of a Pseudo-Random Number Generator (PRNG) needed for the generation of the mask, as well as some additional operations (masking the initial state, unmasking, etc.). Now, if we take a look at the performance data, we can notice that they are almost equal. Indeed, the mask is generated during the key-scheduling, so it does not affect the performances. The slight increase in delay is due to the S-Box operations as well as the masking / unmasking operation.

This table also allows a comparison between the two implementations of Grasshopper and AES-256. First of all, we can see that in terms of space needed on the chip, AES is less expensive than Grasshopper. Actually, the Key-Scheduling method of Grasshopper operates on 128-bit states, whereas for AES the Key expansion uses 32-bit states. Nonetheless, because there are less rounds for Grasshopper (during Key expansion and encryption), the algorithm is faster than AES.

\section{Conducted CPA Attacks}
\label{sec:sec3}

\subsection{Acquisition equipment}

First of all, the targeted FPGA used during these experiments is the Xilinx Artix-7 A35T, embedded on the Basys 3 development board \cite{basys3}. Then, to acquire the electromagnetic leaks, the ChipWhisperer Pro kit, manufactured by NewAE \cite{cwpro}, was chosen.

\begin{center}
  \captionsetup{type=figure}
  \includegraphics[width=16cm]{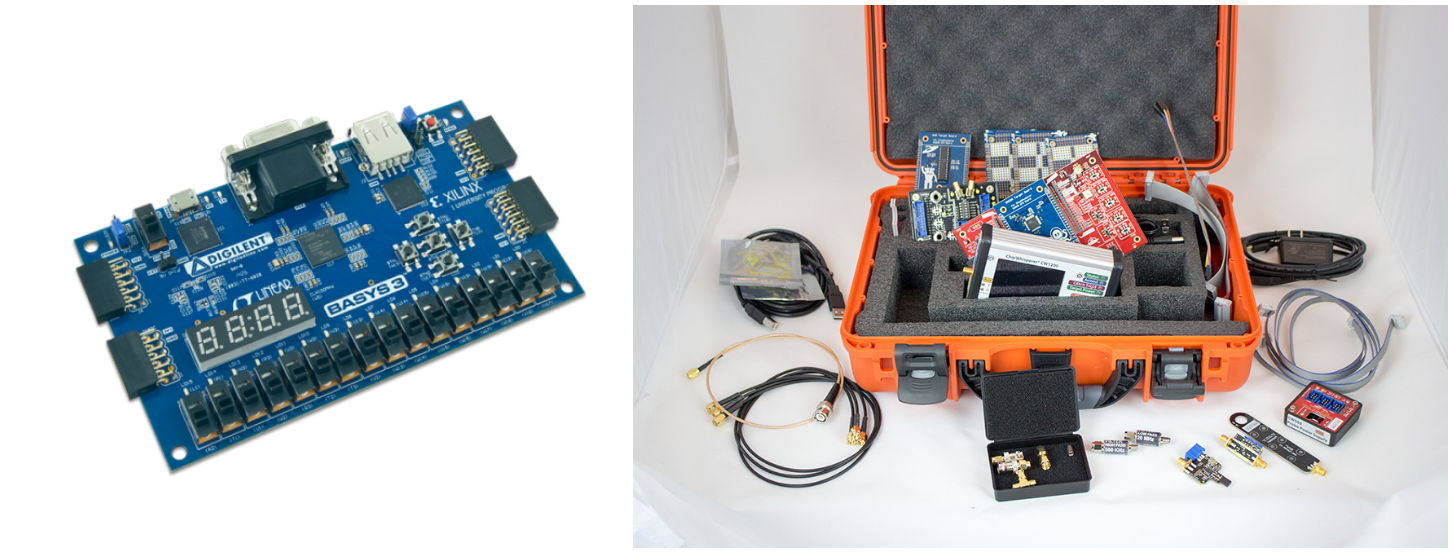}
  \captionof{figure}{Xilinx Basys 3 (on the left) and ChipWhisperer Pro Kit (on the right)}
  \label{fig:fig5}
\end{center}

The setup of trace acquisition can be observed on figure \ref{fig:fig6}. The Basys 3 board is connected to the PC via USB and the plaintexts are sent from the PC to the FPGA through an UART (Universal Asynchronous Receiver Transmitter). The probe that is above the Basys 3 is an H-field probe, that is connected to a Low-Noise Amplifier. This amplifier is then connected to the Chipwhisperer Acquisition equipment which is the ChipWhisperer CW1200 which then transmits the traces to the ChipWhisperer Capture Software thanks to USB connection.

\begin{center}
  \captionsetup{type=figure}
  \includegraphics[width=9cm]{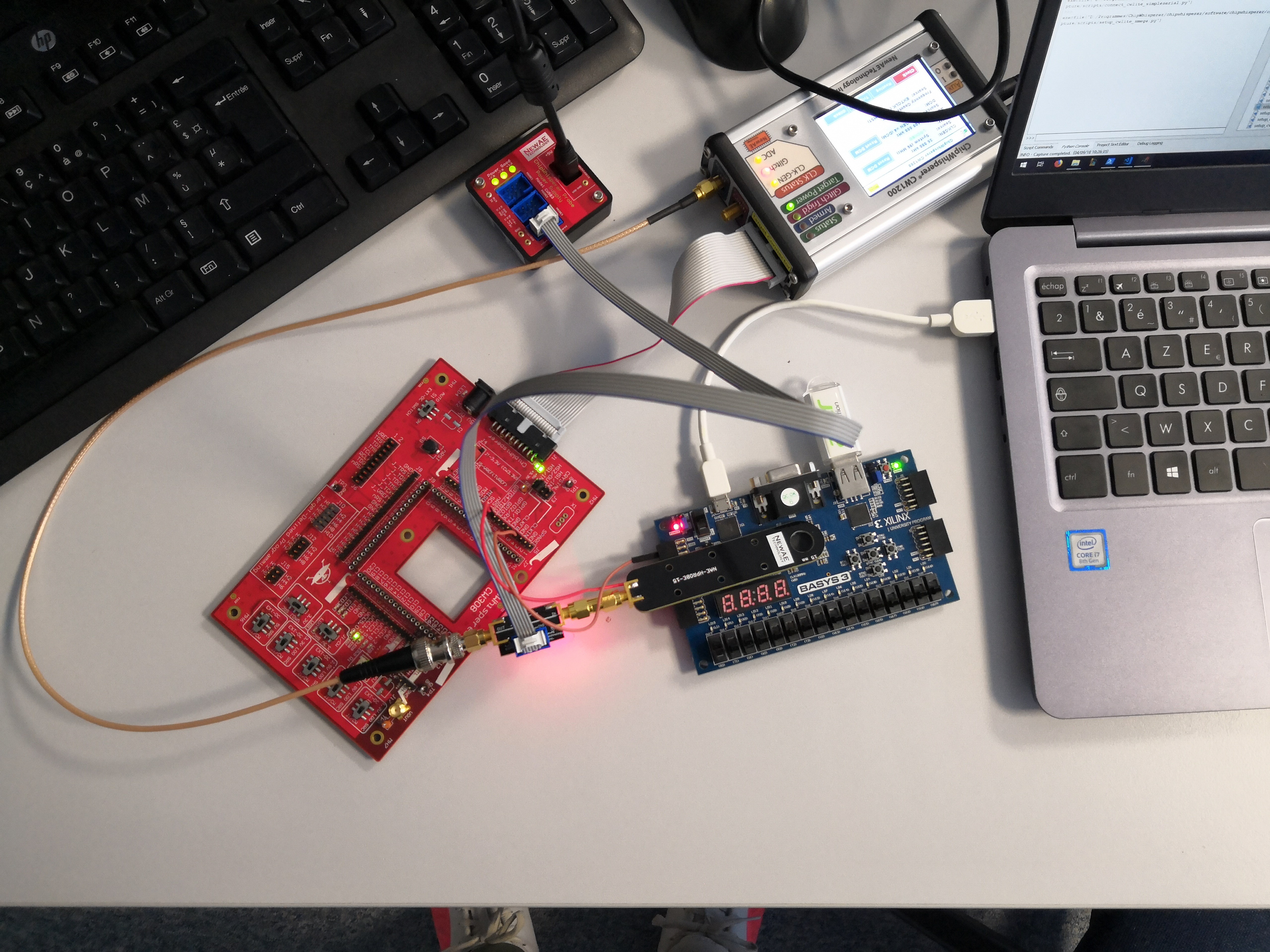}
  \captionof{figure}{Acquisition setup}
  \label{fig:fig6}
\end{center}

Also, in order to be able to acquire these traces, it was necessary to adapt the FPGA design, in particular by creating a UART allowing the transfer of plaintexts and ciphertexts between the PC and the FPGA. The functioning of the FPGA design is therefore presented in Figure \ref{fig:fig7}.

\begin{center}
  \captionsetup{type=figure}
  \includegraphics[width=15cm]{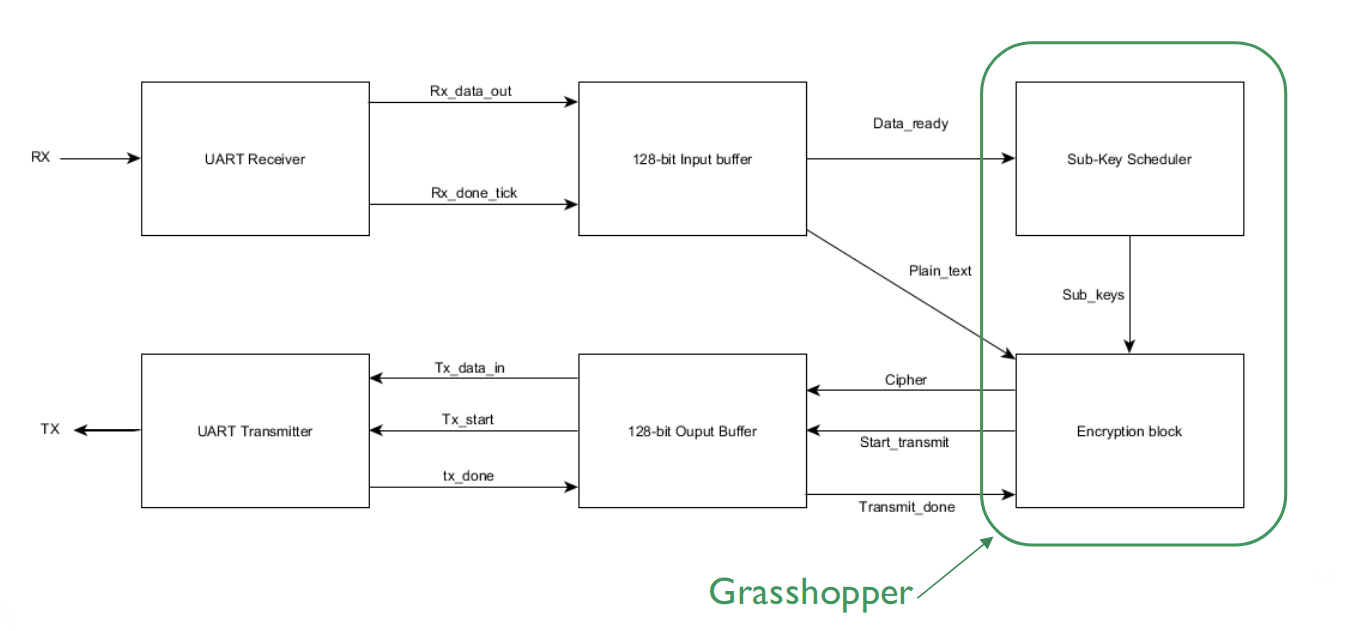}
  \captionof{figure}{Operating diagram of the FPGA design}
  \label{fig:fig7}
\end{center}

The presented setup allows us to acquire the kind of traces presented in Figure \ref{fig:fig8}, which shows a trace acquired during masked encryption, but the standard implementation gives the same trace shapes. In this figure, we can first see the peaks related to the key expansion. Then, the FPGA reads the value of the plaintext before applying the algorithm. We can distinguish the first nine rounds, then the last XOR. The encrypted data is then stored in the buffer that will be read by the UART for transmission to the PC.

\begin{center}
  \captionsetup{type=figure}
  \includegraphics[width=16cm]{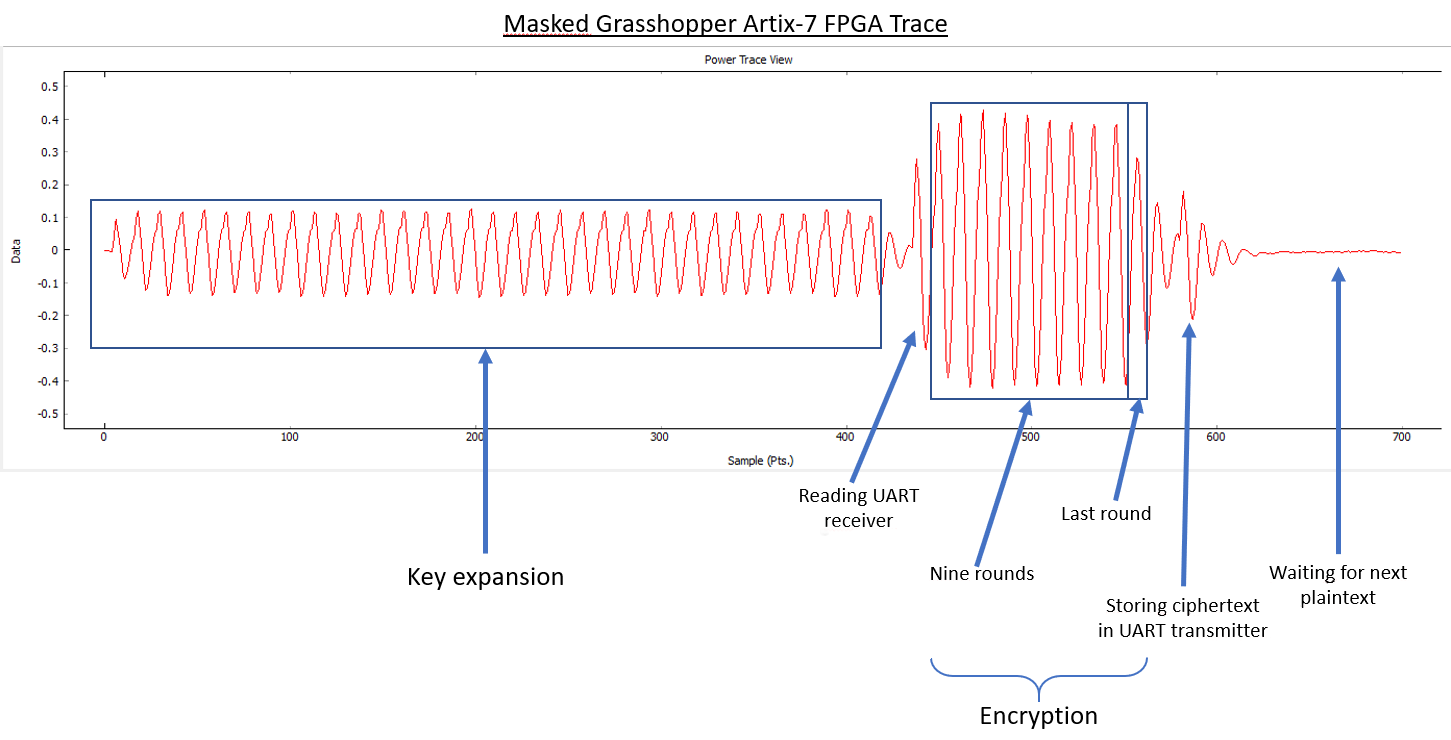}
  \captionof{figure}{Example of Grasshopper trace}
  \label{fig:fig8}
\end{center}

\subsection{Tested attacks and results}

For each of the leakage models studied below, the attacks were carried on a data set composed of 100'000 traces, and the mathematical distinguisher used was the Pearson correlation coefficient, that is defined by the following formula:

\begin{center}
\begin{large}
$ r_{i,j} = \frac{\sum_{d=1}^N\lbrack(h_{d,i}-\bar{h_i})(t_{d,j}-\bar{t_j})\rbrack}{\sqrt{\sum_{d=1}^N(h_{d,i}-\bar{h_i})^{2}\sum_{d=1}^N(t_{d,j}-\bar{t_j})^2}} $
\end{large}
\end{center}

where : 
\begin{itemize}
\item $ i $ is the byte of the tested subkey
\item $ h $ is the value of the theoretical leak (Hamming Distance)
\item $ d $ is the number of the trace
\item $ j $ is the index of the point of the tested trace
\item $ t $ is the value of the trace, at index $ j $
\end{itemize}

In the case of AES, a known CPA attack is that diffusion is not fully ensured on the last lap, since one of the linear functions is not applied on the last lap. Here, the 9 "main" turns of the algorithm are identical, there is simply the last one which has only one XOR with the tenth subkey. So an attack on the last round was tested, with Hamming Distance and Hamming Weight Leakage models, but without any success. Moreover, an attack was intended on the ninth round, supposing that the tenth subkey was known. The attack was also unsuccessful.

\section{Conclusion}
\label{sec:conc}

Two versions of the Grasshopper algorithm have been developed, one optimised but potentially sensitive to side-channel attacks, and the other theoretically resistant to these attacks. However, no known model has been able to fault the algorithm in order to find a subkey.

And even if that had been the case, there is no evidence that the main encryption key could have been found. Indeed, during the generation of subkeys, they are generated in pairs. It is therefore necessary to have these two subkeys to reverse the method of generating subkeys. It is thus necessary to be able to deduce 2 subkeys during the attack. Actually, the subkeys are generated following this formula :
\begin{align*}
    (K_{2i+1}, K_{2i+2}) &= F[C_{8(i-1)+8}]...F[C_{8(i-1)+1}](K_{2i-1}, K_{2i}) \: i = 1,2,3,4 \\
    \intertext{where} F[k](a_1 ,  a_0 ) &= (LSX[k](a_1 ) + a_0 , a_1 )
\end{align*}

Finally, even if the attacks tested were not effective, there is no guarantee that the algorithm implementations are invulnerable to hidden channel attacks. Less "conventional" methods are beginning to appear, particularly based on Deep Learning \cite{deepLearning}, and these methods could prove to be much more effective than the attacks used here.

\section*{Acknowledgements}

I would like to express my sincere thanks to Mr. Alexander A. Istomin (FSUE SIE “GAMMA”), as well as to Mr. Éric Filiol (ESIEA Laval, $(C + V) ^ O $ Lab), for the trust and support they have given me throughout the study.

\bibliographystyle{unsrt}  
\bibliography{references}  

\end{document}